\documentclass[a4paper]{jpconf}
\usepackage{graphicx}

\usepackage{amsmath}
\usepackage{amssymb}
\usepackage{color}

\begin{document}
\title{Recent progress on the accurate determination 
of the equation of state of neutron and nuclear matter}

\author{Paolo Armani, Alexey Yu. Illarionov, Diego Lonardoni, Francesco Pederiva}

\address{Physics Department, University of Trento, and INFN, Gruppo collegato di Trento, via Sommarive, 14, I-38123 Trento, Italy}

\author{Stefano Gandolfi}

\address{Theoretical Division, Los Alamos National Laboratory, Los Alamos, NM 87545, USA} 

\author{Kevin E. Schmidt}

\address{Department of Physics and Astronomy, Arizona State University, Tempe, AZ 85287, USA}

\author{Stefano Fantoni}
\address{ANVUR, National Agency for the Evaluation of Universities and Research Institutes,
Piazzzale Kennedy, 20. I-00144 Roma, Italy}

\ead{pederiva@science.unitn.it}

\begin{abstract}
The problem of accurately determining the equation of state of nuclear and neutron matter at density near and beyond saturation is still an open challenge. In this paper we will review the most recent progress made by means of Quantum Monte Carlo calculations, which are at present the only ab-inito method capable to treat a sufficiently large number of particles to give meaningful estimates depending only on the choice of the nucleon-nucleon interaction. In particular, we will discuss the introduction of density-dependent interactions, the study of the temperature dependence of the equation of state, and the possibility of accurately studying the effect of the onset of hyperons by developing an accurate hyperon-nucleon and hyperon-nucleon-nucleon interaction. 
\end{abstract}

\section{Introduction}

In the last decade the connection between astrophysical observations and the accurate determination of the inter-nucleon forces has become stricter and stricter. This is partly due to the availability of qualitatively much improved data on radii and masses of neutron stars, which now permit, in principle, to exclude some of the models that have been introduced over time. For instance, the recent discovery of very massive neutron star (with estimated mass $> 2M_{\odot}$~\cite{Demorest:2010}), and the general agreement on the fact that maximum masses exceed the value $1.4M_{\odot}$, seem to exclude the hypothesis that at density of order 2$\rho_{0}$, with $\rho_{0}=0.16$ fm$^{-3}$, a significant fraction of hyperons appears, that, according to existing calculations, would give a maximum mass largely below the currently accepted limit~\cite{Vidana:2011,Schulze:2011}. However, our present knowledge of both many-nucleon interaction and of the hyperon-nucleon interaction is too scarce to give to these results any prediction meaning. One of the most important shortcomings of our present knowledge lies in the fact that calculations are performed using approximate interaction in conjunction with approximate methods, and this fact essentially prevents to draw definite conclusions. 

In this paper we will show some of the recent progress that has been made towards an ab-initio treatment of the many-nucleon problem at high densities, by means of Auxiliary Field Diffusion Monte Carlo (AFDMC) and Fermi Hyper-Netted Chain (FHNC) calculations. In particular, we will focus on the recent rediscovery of the idea of modeling many-body interactions by means of density-dependent potential~\cite{Gandolfi:2010,Lovato:2011}, the inclusion of finite temperature effects, and the preliminary work on the inclusion of three-body hyperon-nucleon-nucleon forces in the treatment of a mixed hyperon/nucleon matter. 

\section{Auxiliary Field Diffusion Monte Carlo}

We briefly recall some notions about the AFDMC method. It is part of a wider class of algorithms implementing a stochastic procedure for projecting an initial trial state of an arbitrary Hamiltonian onto the state of lowest energy belonging to the same subspace in the Hilbert space $\{\phi_{n}\}$ of the eigenstates of the Hamiltonian itself: 
\begin{equation}
\lim_{\tau\rightarrow\infty}\Psi^{\alpha}(R,\tau) = \lim_{\tau\rightarrow\infty}\sum_{n}c_{n}e^{-\tau(\hat{H}-E^{\alpha}_{0})}\phi_{n}=c^{\alpha}_{0}\phi^{\alpha}_{0}.
\end{equation}
The index $\alpha$ represents a specific symmetry for the state described by a set of quantum numbers, and/or the symmetry imposed by the Pauli principle. The constant $E_{0}^{\alpha}$ is a reference energy needed to control the normalization of the propagated states. 
The propagation is obtained by expanding the initial state on a finite number eigenstates of the position and spin/isospin operators $\{|x\rangle\}$. Each initial point is then propagated by sampling a kernel, which, in general, is approximated by a Trotter-Suzuki break up of the original propagator for small imaginary time-steps:
\begin{equation}
\langle x'|e^{-\Delta\tau\hat{H}}|x\rangle\langle x|\Psi\rangle=\langle x'|e^{-\Delta\tau[\hat{V}-E_{0}^{\alpha}]}|x'\rangle\langle x'|e^{-\Delta\tau\hat{T}}|x\rangle\langle x|\Psi\rangle +o(\Delta\tau).
\end{equation}
The resulting expression for the propagation of a set of configurations of A particles can be summarized in integral form:
\begin{equation}
\Psi(R',\Delta\tau) = \left(\sqrt{\frac{m}{2\pi\hbar^{2}\Delta\tau}}\right)^{3A}\int\; dR \exp[-\Delta\tau(V(R')-E_{0}^{\alpha})]\exp\left[-\frac{(R-R')^{2}}{2\hbar^{2}/m\Delta\tau}\right]\Psi(R,0).
\end{equation}
Therefore, the state is propagated by sampling the propagator and applying it to a set of points representative of the initial wave function. The points are iteratively displaced and weighted according to the various terms in the propagator, until a sufficiently long imaginary time is reached. In general the distribution of points is importance sampled, in the sense that it is multiplied by some approximate expression of the wave function of the state searched in order to avoid the large fluctuations in the normalization coming from the possibly divergent behavior of the potential energy. The scheme works flawlessly only if the projected state is the absolute ground state of $\hat{H}$, i.e., if it is a function symmetric under particle exchange. For a many Fermion system, as in the case of nucleons, the calculation is proved to be unstable, in that the variance/average ratio diverges exponentially. In this case it is customary to apply some constraint on the sampled configurations in order to maintain the overlap between the sampled distribution of points and the state onto which we want to project. These procedure are approximate, because they introduce a (usually small) bias on the estimates by effectively changing the Hamiltonian used in the propagation.

An additional source of difficulty is introduced by the fact that, for many-nucleon systems, the potential is usually non-local. Local expressions of the two-body interaction, like the Argonne AV18 potential, are written as a sum over operatorial components:
\begin{equation}
V(r_{ij})=\sum_{p=1}^{18}v_{p}(r_{ij})O_{p}.
\end{equation}
For the simple AV6 case, discussed in this paper, the operators considered are:
\begin{equation}
O_{p}=(1,\tau_{i}\cdot\tau_{j})\otimes(1,\sigma_{i}\cdot\sigma_{j},S_{ij}),
\end{equation}
where $S_{ij}$ is the tensor operator. From the point of view of the Diffusion Monte Carlo (DMC) algorithm, the use of such potentials imply that the factor $ \exp[-\Delta\tau(V(R')-E_{0}^{\alpha})]$ cannot be treated as a local ``weight'' for the propagated points, but rather behaves as an imaginary time propagator for the spin/isospin degrees of freedom.  In the AFDMC algorithm the problem is circumvented by recasting the interaction in a spin/isospin independent part $V_{si}$ and a spin/isospin dependent one $V_{sd}$. The second is in turn written as a bilinear form in the spin/isospin operators. For simplicity, let us drop the isospin components. The interaction becomes:
\begin{equation}
V= V_{si}+\frac{1}{2}\sum_{i,\gamma,j,\delta}\sigma_{i,\gamma}A_{i,\gamma,j,\delta}\sigma_{j,\delta}.
\end{equation}
By diagonalizing the matrix $A$, and indicating as $\varphi_{n}$ and $\lambda_{n}$ the corresponding eigenvectors and eigenvalues, we have:
\begin{equation}
V=V_{si}+\frac{1}{2}\sum_{n=1}^{3A}\lambda_{n}O_{n}^{2},
\end{equation}
where:
\begin{equation}
O_{n}=\sum_{j,\delta}\varphi_{n}^{j,\delta}\sigma_{j,\delta} .
\end{equation}
At this point, in the small $\Delta\tau$ limit, the spin dependent propagator can be recast by using the Hubbard-Stratonovich transformation in order to reduce the operatorial dependence from quadratic to linear:
\begin{equation}
e^{-\Delta\tau \hat{V}_{sd}}\simeq\prod_{n=1}^{3A}e^{-\frac{1}{2}\lambda_{n}O_{n}^{2}\Delta\tau}=\prod_{n=1}^{3A}\frac{1}{\sqrt{2\pi}}\int dx_{n}e^{-\frac{x_n^{2}}{2}}e^{x_{n}\sqrt{-\lambda_{n}\Delta\tau}O_{n}}.
\end{equation}
The main advantage of this procedure is in avoiding the necessity of a sum over all the components of the wave function corresponding to different two-nucleon states. This number grows exponentially with the number of nucleons, and prevents the straightforward application of Green's Function Monte Carlo (GFMC) techniques for systems larger than $A=12$~\cite{Pieper:2007}. However, the price to pay is the introduction of a set of auxiliary degrees of freedom by means of which rotation of the spinorial components of the single particle functions are sampled. Treating a sufficiently large number of nucleons in a periodic box is a necessary condition for studying the properties of infinite matter without suffering too heavy finite-size effects. A study of the convergence of the results to the thermodynamic limit can be found in Refs.~\cite{Sarsa:2003,Gandolfi:2009}.

The AFDMC algorithm actually in use exploits a number of technicalities that are not reported here, including an extension of the importance sampling concept, and a particular form of constraint to avoid the problem of the exponential growth of the variance. All such details have been largely discussed in the existing literature.
For neutron matter and neutron drops the method can be  extended
to Hamiltonians including spin-orbit (AV8') and three-body
interactions~\cite{Pederiva:2004,Gandolfi:2011,Gandolfi:2011b}. In the case of nuclear
matter the method is still limited to the use of the 6 operators mentioned
above~\cite{Gandolfi:2007}.

\section{Density Dependent Interactions}

An accurate description of the many-nucleon interaction in nuclear matter should in principle rely on the forces that describe the binding energies of nuclei. However, at present, the simple transposition of the Hamiltonian as is from nuclei to the infinite systems seems not to give satisfactory results. This is obviously due to the difficulty of developing a true ab-initio theory of the many-nucleon interaction. 
In order to be operative, and provide to astrophysicists some significant estimates of the equation of state, it is possible to implement a two-body potential that effectively includes many-body effects through a density dependence of the coupling constants.
Following Lagaris and Pandharipande, the  AV6 potential, can be recast by introducing a density dependent intermediate part:\begin{equation}
v_{ij}=v_{\pi}+e^{-\gamma_{1}\rho}v_{I}+v_{r}.
\end{equation}
In order to reproduce the saturation density and energy, it is then necessary to introduce a purely phenomenological attractive contribution which includes a symmetry term:
\begin{equation}
TNA = \gamma_{2}\rho^{2}e^{-\gamma_{3}\rho}\left[3-2\left( \frac{\rho_{n}-\rho_{p}}{\rho}\right)\right].x
\end{equation}
\begin{figure}[t]
\begin{center}
\includegraphics[scale=0.6]{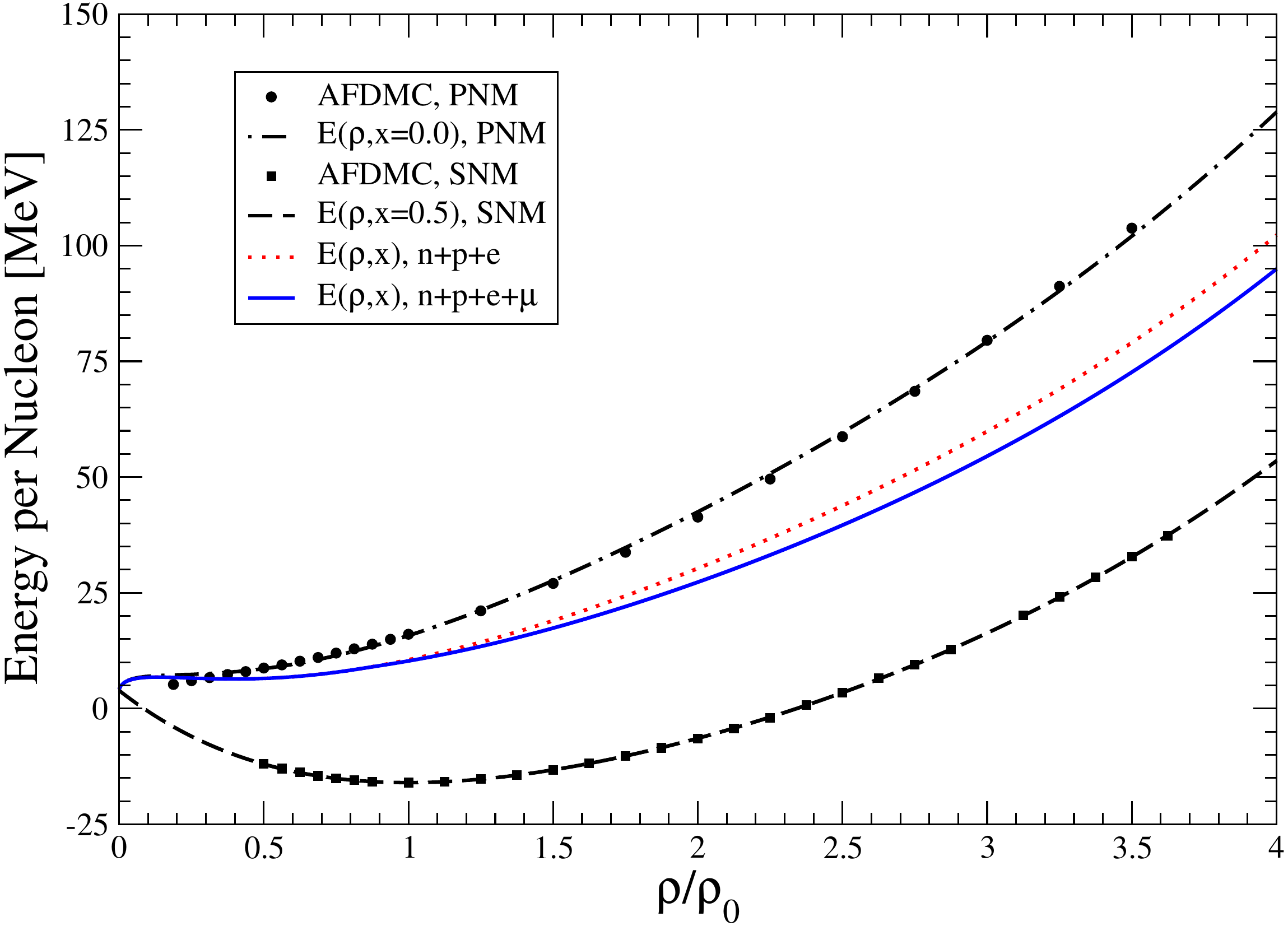}
\end{center}
\caption{\label{label}Equation of state of symmetric and neutron matter from AFDMC calculations with a density dependent interaction (see legend). The curve computed at $\beta$-equilibrium with and without inclusion of muons is also reported. }
\end{figure}

The three parameters are fitted, by means of AFDMC calculations, towards the saturation energy, saturation density and compressibility in nuclear matter~\cite{Gandolfi:2010}. Their values are $\gamma_{1}=0.10$ fm$^{3}$, $\gamma_{2}=-750$ MeV$\cdot$fm$^{6}$, and $\gamma_{3}=13.9$ fm$^{3}$ respectively. With these values is then possible to evaluate the energy per nucleon at densities above and below saturation. The resulting equation of state of the Symmetric Nuclear Matter (SNM) can be fitted by the expression:
\begin{equation}
\frac{E_{SNM}(\rho)}{N}=E_{0}+b(\rho-\rho_{0})^{2}+c(\rho-\rho_{0})^{3}e^{\gamma(\rho-\rho_{0})},
\end{equation}
with $E_{0}=-16.0$ MeV, $\rho_{0}=0.16$ fm$^{-3}$, $b=520.0$ MeV$\cdot$fm$^{6}$, $c=-1297.4$MeV$\cdot$fm$^{9}$, and $\gamma=-2.213$fm$^{3}$~\cite{Gandolfi:2010}. In Fig. 1 we report the corresponding curve, together with the results of the equation of state of pure neutron matter (PNM) obtained computing the energy per neutron with the same density dependent interaction (DDI).  Assuming a quadratic behavior in the proton/neutron number difference for the symmetry energy, it is possible to write an equation of state for a generic proton fraction $x_{p}$ as follows:
\begin{equation}
\label{eq:esym}
E(\rho,x_{p})=E_{SNM}(\rho)+C_{s}\left(\frac{\rho}{\rho_{0}}\right)^{\gamma_{s}}(1-2x_{p})^{2},
\end{equation}
with $C_{s}=31.3$MeV, and $\gamma_{s}=0.64$.
Typical values for these parameters have been quoted as $C_s \approx 31-33$ MeV and
$\gamma_s \approx 0.55-0.69$ by \cite{Shetty:2007} and as $C_s = 31.6$ MeV and
$\gamma_s \approx 0.69-1.05$ by \cite{Worley:2008}.
It should be noticed that, usually, the symmetry energy is constrained over a range of densities
typical of nuclei, whereas here it was fitted the parameters over a very wide density range.
This means that the parametrization of Eq.~(\ref{eq:esym}) should be accurate up to very high densities.

 In this way it is possible, density by density, to compute the equation of state at beta-equilibrium with electrons and muons, by assuming that the chemical potential of the leptons is described by that of an ultra-relativistic Fermi gas. The resulting curves are also reported in Fig. 1. 

\section{Finite Temperature Corrections}

In the final stages of a supernova explosion, the dynamics of the resulting proto neutron star is driven by the equation of state at temperatures that initially reach values of about 20 MeV. In astrophysical models the temperature dependence of the equation of state is usually assumed to be very simple. So far, realistic calculations have been performed in the Brueckner-Hartree-Fock framework\cite{Nicotra:2006}.
In order to have a completely ab-initio description it would be necessary to turn to Path Integral based Monte Carlo methods that allow to compute expectations over the quantum thermal matrix. However, this step implies a number of technical difficulties that have not been faced yet. 

An interesting intermediate approach consists of estimating the thermal corrections to the zero temperature EoS by means of of a temperature dependent FHNC calculation\cite{Schmidt:1979}. Such corrections can then be applied to the equation of state computed by AFDMC to obtain a result as accurate as possible.
Several tests showed that by including the temperature effects using FHNC the contributions of the most important 
elementary diagrams cancel, and the difference is only weakly affected by the lack of such diagrams.

By extending the variational chain summation to finite temperatures one has to face the so-called
orthogonality corrections \cite{Schmidt:1979,Fantoni:1988}, since wave functions used are not mutually orthogonal.
The orthogonality corrections are not unique and, in general, their calculation requires
the evaluation of the off-diagonal matrix elements of the Hamiltonian.
At present no accurate method exists to efficiently compute such off-diagonal matrix elements.
Recently it was proved that the orthogonality corrections to the free energy vanish in the
thermodynamic limit \cite{Mukherjee:2007}.
In the spirit of Landau's theory of Fermi liquids,
where it is assumed that the low-lying eigenstates of an interacting system have one-to-one correspondence with those
of the noninteracting gas,
we assume that a good approximation for the eigenstates of the interacting fermion system is given
by the correlated basis states \cite{Schmidt:1979}:
\begin{equation}
	\Psi_i\left[n_i(\mathbf{k})\right] =
	\frac{\mathcal{S}\left(\prod_{i<j}\mathcal{F}_{ij}\right)\Phi_i\left[n_i(\mathbf{k})\right]}
	{\sqrt{\Phi_i^\dag\left[n_i(\mathbf{k})\right] \left(\mathcal{S}\left(\prod_{i<j}\mathcal{F}_{ij}\right)\right)^2
	\Phi_i\left[n_i(\mathbf{k})\right]}} \,,
\label{CBS}
\end{equation}
where $\Phi_i$ are the single particle states of the non-interacting system . The $n_i(\mathbf{k})=0,1$ are the
occupation numbers for single particle states labeled by $\mathbf{k}$.
The pair correlation operator defining the correlated wave function $\mathcal{F}_{ij}$  is taken to be
\begin{equation}
	\mathcal{F}_{ij}=\sum_{p=1}^6 f_p(r_{ij}) O_{ij}^p \,,
\label{corr.op}
\end{equation}
where $p=1-6$ operators are defined as in Eq. (5).
Since the operators $O_{ij}^p$ do not commute, the product of correlation operators is symmetrized
with the symmetrization operator $\mathcal{S}$ to make the wave function antisymmetric.

The upper bound for the free energy $F(\rho,T)$ can be obtained by using the Gibbs-Bogoliubov variational
principle
\begin{equation}
	F(\rho,T) \le F_V(\rho,T) = {\rm Tr}\left(\rho_V H\right) - TS_V(\rho,T) \,.
\label{var:FT}
\end{equation}
$\rho_V$ is any arbitrary density matrix satisfying:
\begin{equation}
	{\rm Tr}\rho_V = 1,
\label{norm:rhoV}
\end{equation}
and $S_V(\rho,T)={\rm Tr}\left(\rho_V\ln\rho_V\right)$ is the entropy derived from the density matrix $\rho_V$ at temperature $T$.
The equality in (\ref{var:FT}) holds when $\rho_V$ is the exact density matrix of the system.
In order to obtain $\rho_V$ it is necessary to start from an ansatz for a correlated effective
Hamiltonian $H_V$, such that:
\begin{equation}
	\rho_V=\frac{\exp\left(-\beta H_V\right)}{{\rm Tr}\exp\left(-\beta H_V\right)} \,,
\label{rhoV:HV}
\end{equation}
where $\beta=1/T$ is the inverse temperature ( $k_B=1$).

In practice $H_V$ is chosen to be a one-body operator such that correlated basis states (\ref{CBS}) are eigenstates of it:
\begin{equation}
	H_V|\Psi_i\left[n_i(\mathbf{k})\right]\rangle =
	\left[\sum_\mathbf{k} n_i(\mathbf{k})\epsilon_V(\mathbf{k},\rho,T)\right] |\Psi_i\left[n_i(\mathbf{k})\right]\rangle \,.
\label{HV:psi}
\end{equation}
The eigenvalues of this $H_V$ can be varied by changing the single-particle spectrum $\epsilon_V(\mathbf{k})$,
which can be interpreted as the quasi-particle spectrum, and the eigenfunctions by varying the correlation operator $\mathcal{F}_{ij}$.

With this choice of $H_V$, the calculation of the entropy $S_V$ is trivial.
At temperature $T$ the average occupation number of a single-particle state is given by:
\begin{equation}
	\bar{n}(\mathbf{k},\rho,T) = \frac{1}{\exp\bigl[\beta\bigl(\epsilon(\mathbf{k},\rho,T)-\mu(\rho,T)\bigr)\bigr]+1} \,,
\label{n:aver}
\end{equation}
where the chemical potential $\mu(\rho,T)$ is required to satisfy 
the normalization condition:
\begin{equation}
	A=\sum_\mathbf{k} \bar{n}(\mathbf{k},\rho,T) \,,
\label{def:mu}
\end{equation}
where $A$ is  the total number of particles in the system,
and the entropy is given by:
\begin{eqnarray}
	S_V(\rho,T) = -\sum_\mathbf{k} \Bigl[\bar{n}(\mathbf{k},\rho,T)\ln\left(\bar{n}(\mathbf{k},\rho,T)\right)
	+ \left(1-\bar{n}(\mathbf{k},\rho,T)\right)\ln\left(1-\bar{n}(\mathbf{k},\rho,T)\right)\Bigr] \,.
\label{SV}
\end{eqnarray}

Since the correlated basis states (\ref{CBS}) are not mutually orthogonal, the last equation is only
an approximation if the variational Hamiltonian $H_V$ is defined by Eq.~(\ref{HV:psi}).
It is exact if only orthonormalized correlated basis states are be used.

The calculation of $E_V(\rho,T)/A={\rm Tr}\left(\rho_V H\right)$ is very similar to the variational calculation of the
ground state expectation value $E_0$ by expanding it in power of $\mathcal{F}_{ij}^2-1$.
Schematically, we have:
\begin{equation}
	\frac{E_V(\rho,T)}{A}=\frac{\hbar^2k_{\rm av}^2}{2m} + \sum{\rm diagrams}\bigl(v,\mathcal{F},\ell(r,\rho,T)\bigr) \,,
\label{EV}
\end{equation}
where $m$ is the average bare mass of a nucleon and $k_{\rm av}^2$ is a mean square momentum per particle,
\begin{equation}
	k_{\rm av}^2 = \frac{1}{A} \sum_\mathbf{k} \mathbf{k}^2 \bar{n}(\mathbf{k},\rho,T) \,.
\label{kav}
\end{equation}
The diagrams are many-body integrals involving the potential $v$, the correlation operator $\mathcal{F}$ and
the finite temperature Slater function $\ell$,
\begin{equation}
	\ell(r,\rho,T) = \frac{1}{A} \sum_\mathbf{k} \exp(i\mathbf{kr}) \bar{n}(\mathbf{k},\rho,T) \,.
\label{ell}
\end{equation}
\begin{figure}[t]
\begin{center}
\includegraphics[width=0.7\textwidth,scale=1.0]{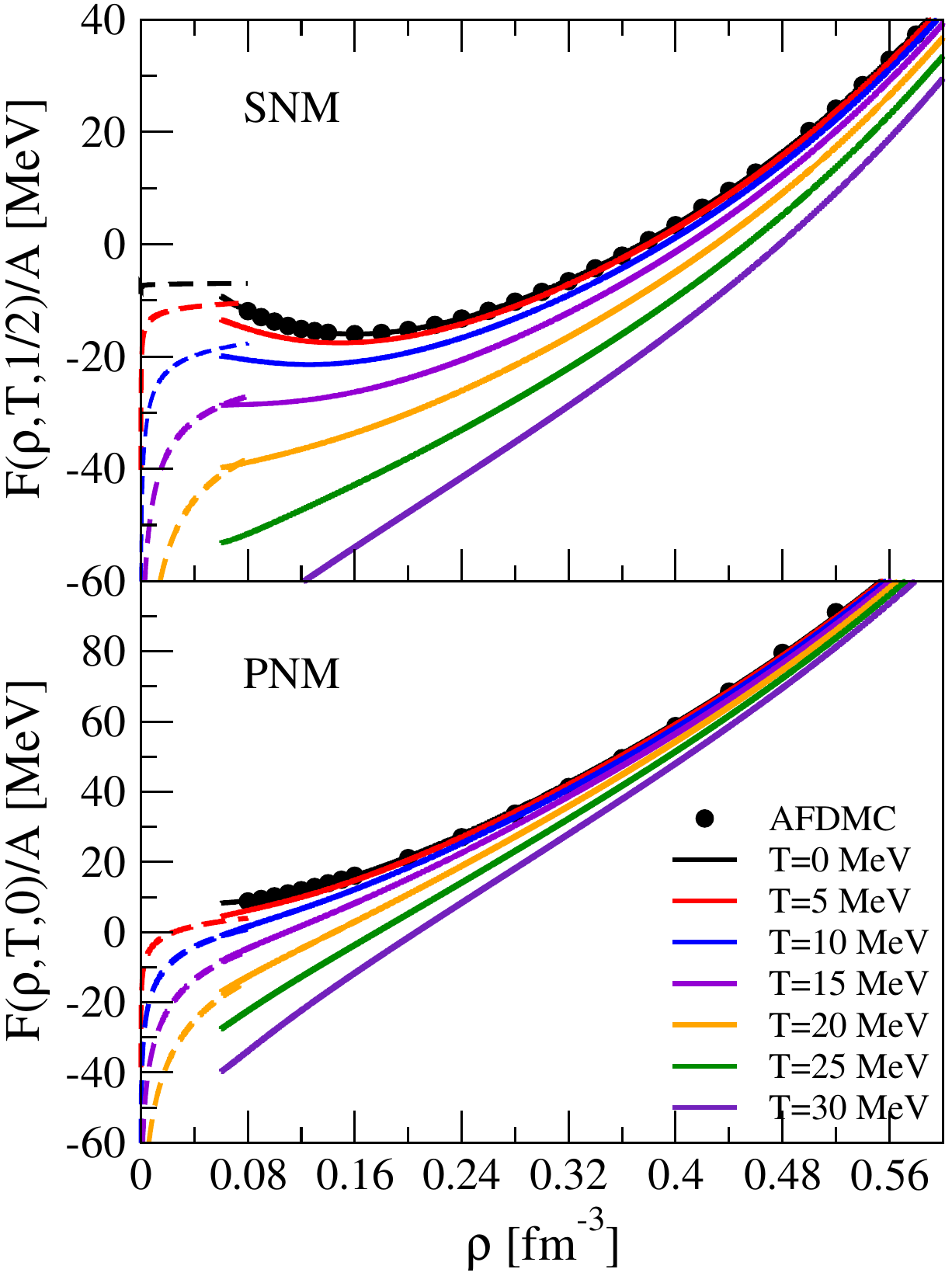}
\end{center}
\caption{Equation of state of Pure Neutron Matter (bottom panel) and Symmetric Nuclear Matter (top panel), computed by adding the estimate of the FHNC free energy estimation to the zero temperature results obtained by means of AFDMC calculations. Results are reported for temperatures ranging from 0 to 30 MeV. The low density PNM EoS is fitted to the virial equation of state of Ref.~\cite{Shen:2010,Horowitz:2006}}
\end{figure}

The single particle spectrum is parametrized as:
\begin{equation}
	\epsilon(\mathbf{k},\rho,T) = \frac{\hbar^2k^2}{2m\Bigl[1+A(\rho,T)\exp\bigl(-B(\rho,T)k^2\bigr)\Bigr]} \,.
\label{e:param}
\end{equation}
It should be noted that
in general it is possible for $\epsilon(\mathbf{k},\rho,T)$ to have higher order terms in $k$.
However, the free energy was found to be insensitive to any such dependence at moderate density range.
Under this assumption, if $B(\rho,T)=0$, the spectrum is fully determined by an effective mass:
\begin{equation}
	\frac{m^\star(\rho,T)}{m}=
	\frac{\hbar^2}{m}\left(\frac{1}{k}\frac{d\epsilon}{dk}\right)^{-1} =
	1+A(\rho,T) \,,
\label{eff.mass}
\end{equation}
that can vary with temperature and density.

This scheme has been applied to compute the thermal corrections to SNM and PNM within the FHNC/SOC scheme. The EoS resulting by adding such corrections to the AFDMC result obtained with a DDI  is plotted in Fig. 2.
The following functional form provides a good parametrization of the numerical results in the required ranges of density and temperature:
\begin{eqnarray}
	F(\rho,T,x)/A = E(\rho,x)/A 	+ \Delta F_0(\rho,T)/A + (1-2x)^2 \Delta F_S(\rho,T)/A\\\nonumber
	- \alpha \left(\frac{\rho_0}{\rho}\right)^\beta \left[x^{1/3}+(1-x)^{1/3}\right] T^2 \,,
\label{eq:eos-T}
\end{eqnarray}
where $\alpha$ and $\beta$ are almost independent on the isospin $x$. The fit is inspired by the Sommerfeld expansion, and resembles the excitation energy of a hot non-interacting Fermi gas \cite{Huang:1963}:
\begin{eqnarray}
	\left(F-F_0\right)/A
		&= - \frac{3\pi^2}{8\mu_F}(k_BT)^2 \left[x^{1/3}+(1-x)^{1/3}\right] + O(T^4)\\\nonumber
		&\approx -a_e\left(\frac{\rho_0}{\rho}\right)^{2/3} \left[x^{1/3}+(1-x)^{1/3}\right] T^2 \,.
\label{eq:exT}
\end{eqnarray}
For pure neutron matter at the normal density the parameter $a_e$ has the value $3\pi^2/(8\mu_F)=0.03315~{\rm MeV}^{-1}$.

Other functions, entering the definition of (\ref{eq:eos-T}), are the following:
\begin{eqnarray}
	\Delta F_0(\rho,T)/A
		&=& \left[a_T \log\rho + b_T \left(\frac{\rho_0}{\rho}\right)\right] T
		+ \left[c_T \log^2\rho + d_T \left(\frac{\rho_0}{\rho}\right)\right] T^2 \,,
\label{eq:eos-F_0} \\
	\Delta F_S(\rho,T)/A
		&=& e_T \left(\frac{\rho_0}{\rho}\right) T^2 \,.
\label{eq:eos-F_S}
\end{eqnarray}

The parameters have been fitted against the AFDMC results. Their values are the following:
$a_T=-0.15(2)$, $b_T=-0.38(4)$, $c_T=-0.008(1)$, $d_T=0.06(3)$, $e_T=-0.016(13)$,
$\alpha=0.047(23)$, $\beta=0.72(14)$ with $\chi^2/n.d.f=0.54$.

The entropy per nucleon $S(\rho,T,x)$, which is a measure of thermal disorder,
is calculated from the quasi-particle occupation probabilities
$\bar{n}(\mathbf{k},\rho,T)$ using Eq.~(\ref{SV}).
It is also possible to compute $S(\rho,T,x)$ by means of the following expression:
\begin{equation}
	S(\rho,T,x) = -\left(\frac{\partial F/A}{\partial T}\right)_V \,.
\label{F->S}
\end{equation}
The values of the entropy computed by the two different procedures are in excellent agreement. This is a
strong test of the quality of the variational calculation of  $F(\rho,T,x)$
as discussed in \cite{Friedman:1981}.
\begin{figure}[t]
\begin{center}
\includegraphics[width=10cm]{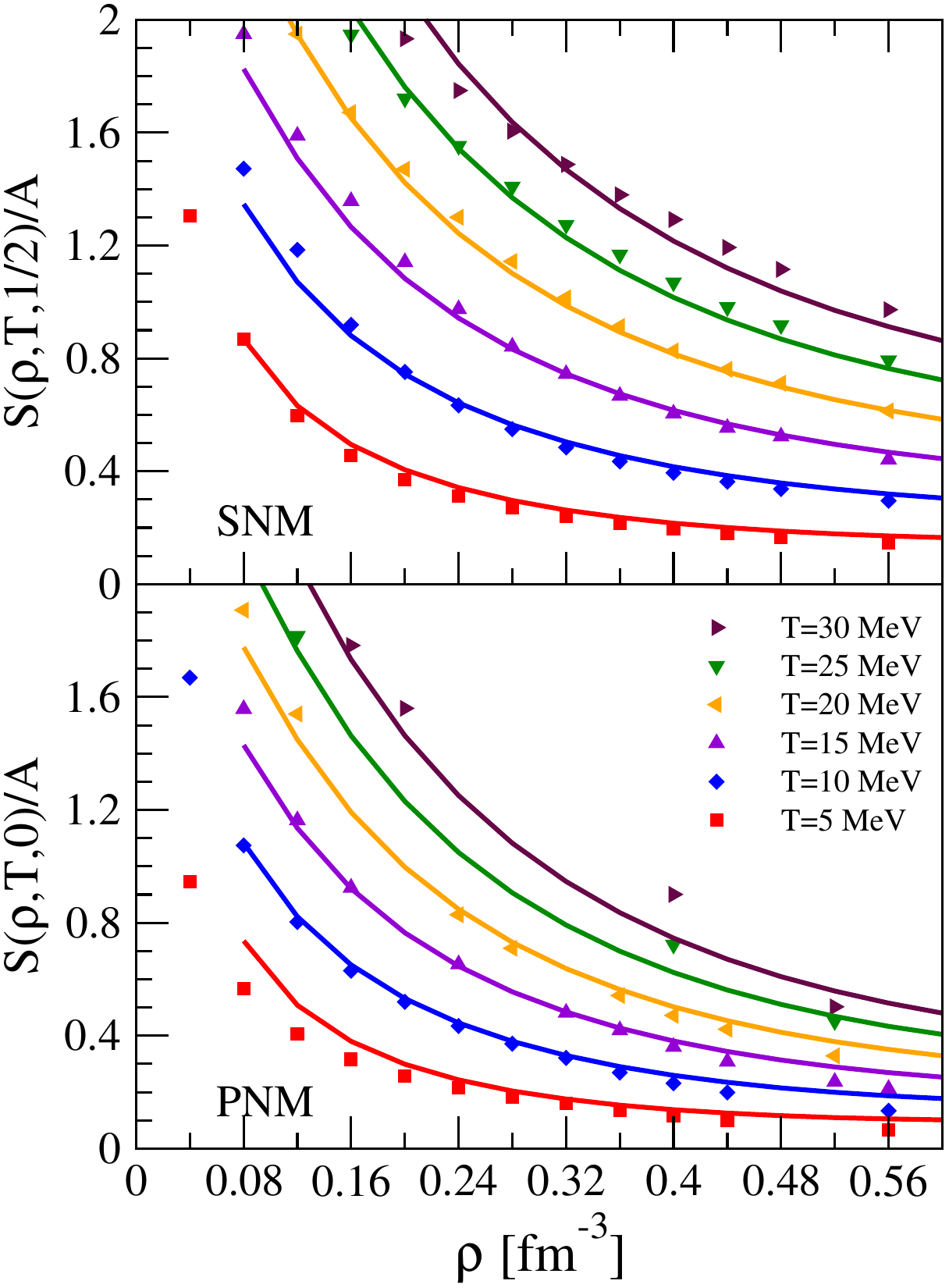}
\end{center}
\caption{\label{fig:entropy-T}Entropy per nucleon in SNM (upper panel) and PNM (lower panel) as a function of the density of the system. }
\end{figure}
Notice that entropy production in multi-fragmentation events in heavy-ion collisions
is a crucial quantity in the determining the mass fragment distribution.
The entropy per particle is shown in Fig.~\ref{fig:entropy-T} as a function of density and for various temperatures.
The entropy increases with temperature, as physically reasonable, and decreases substantially with density.
At low $T$, it is expected to approach a linear dependence due to the fact that, for a Fermi liquid,
the relation between $S$ and $T$ should be approximately
\begin{equation}
	S \approx \frac{\pi^2}{3\rho} N(T=0) T = \frac{\pi^2 m^\star}{\hbar^2k_F^2} T \,,
\label{entropy-appr}
\end{equation}
in terms of the density of states at the Fermi surface. The specific heat is defined as
\begin{equation}
	C_V(\rho,T,x) = T \left(\frac{\partial S}{\partial T}\right)_V \,.
\label{S->cv}
\end{equation}

The energy density $\epsilon(\rho,T,x)$, sound velocity $c_s(\rho,T,x)$ (in units of $c$)
and adiabatic index $\Gamma(\rho,T,x)$ are given by:
\begin{eqnarray}
	\epsilon(\rho,T,x)&=&\rho\left(F(\rho,T,x)/A+(1-x)m_nc^2+x m_pc^2\right) \,,
\label{en.dens.}\\
	c_s(\rho,T,x)&=&\sqrt{\frac{\partial P(\epsilon)}{\partial\epsilon}} \,,
\label{sound}\\
	\Gamma(\rho,T,x)&=&\frac{\epsilon}{P} c_s^2 \,.
\label{a.index}
\end{eqnarray}
If $\Gamma$ is constant, then the EOS becomes of the usual polytrope form, $P \sim \epsilon^\Gamma$.

\section{Hyperons and hypernuclei}

The onset of degrees of freedom with strangeness $S\neq 0$ in nuclei and nuclear matter is a problem of 
great interest in the study of the properties of dense stars. At densities of about $2\rho_{0}$, where
$\rho_{0}=0.16$ fm$^{-3}$ is the saturation density of the nuclear  medium, the chemical potential of
the ultra-relativistic electron gas, determined by the $\beta$-equilibrium condition becomes comparable with 
that of the $\Sigma^{-}$ hyperon, which becomes stable due to its larger mass and the consequent 
decrease in kinetic energy.
This fact has strong consequences on the EoS, which becomes softer than that predicted
by models in which strange degrees of freedom are absent. Present calculations of the equation of 
state (EOS) of dense matter including hyperons show that this softening leads to an unphysical limitation
of the maximum observable mass of a star~\cite{Vidana:2011,Schulze:2011}. This seems to be an indication against the indirect
evidence of the occurrence of this mechanism.
On the other hand, most calculations neglect important pieces of the interaction, and in particular 
the hyperon-nucleon-nucleon (YNN) contribution, which, if repulsive on average, might in principle completely
change this picture. Due to the fact that the YN interaction must be mediated by at least two pions, the YNN force
appears at the same order, and cannot be assumed to be small.

Recently we started an ambitious project, which should lead to a more accurate determination of the YN and
YNN interactions combining AFDMC calculations and possibly new available data on separation energies in hyper nuclei. 
Due to the very limited availability of data on $\Sigma$-hypernuclei, it is necessary to focus on the $\Lambda$N and
$\Lambda$NN interactions only.
The starting model is that introduced by Usmani \cite{Usmani:1995}.
The system under study is a hypernucleus composed by $A$ nucleons interacting through a two body AV6' potential, and one hyperon.
We write the Hamiltonian of the system as:
\begin{eqnarray}
	H_{N+\Lambda}&=&H_N+H_\Lambda= 
	-\sum_{i=1}^A\frac{\hbar^2}{2m_i}\nabla_i^2 + \sum_{j>i=1}^A V^{NN}(i,j)
	-\frac{\hbar^2}{2m_\Lambda}\nabla_\Lambda^2 + \sum_{i=1}^A V^{\Lambda N}(\Lambda,i) + \ldots
\end{eqnarray}
The explicit form of the  $\Lambda$-nucleon potential is even in
Ref. \cite{Usmani:1995}. It essentially accounts for a two-pion exchange
interaction. In principle, it should also include a contribution from
Kaon exchange. This term contributes to the tensor components, but it
also includes a space exchange term between the nucleon and hyperon
degrees of freedom. As assumed, and partially justified, in other
works like \cite{Usmani:O17,Usmani:1999}, we  assume that the
exchange term is quite negligible. The contribution can be considered
as effectively included in the hyperon separation energy by the fitted
value of the strength of the two--pion exchange term.

The three-body YNN interaction is instead of the general form:
\begin{equation}
V_{YNN}=V^{D}+V^{2\pi},
\end{equation}
where $V^{D}$ is a dispersive term, an $V^{2\pi}$ is once more a standard two-pion exchange contribution.
It does not present any special difficulty in the AFDMC scheme, due to the assumed distinguishability of the $\Lambda$ with respect to the nucleons. Particular care has to be taken in treating the center of mass contributions.

The separation energy of the $\Lambda$ particle is defined starting from the difference between the energies of the
nuclear systems with and without the $\Lambda$ hyperon:
\begin{equation}
	-B(\Lambda)= \langle H_{N+\Lambda}\rangle_{A+\Lambda} -
		\langle H_N \rangle_A \,,
\label{def:BL}
\end{equation}
The hypernucleus wave function is built starting from single particle orbitals computed by HF with a Skyrme I force. 
A general expression of the mean-field part of the wave function is then:
\begin{equation}
\label{psil}
	|\Psi_{A+\Lambda}\rangle = \left( \prod_{i=1}^A f_\Lambda (r _{\Lambda i}) \right)
	\phi_{nljm}^\Lambda ({\bf r}_\Lambda)|\Psi_{A}\rangle \,,
\end{equation}
where $\phi_{nljm}^\Lambda ({\bf r}_\Lambda)$ is the orbital describing the hyperon.
The function $f_\Lambda (r _{\Lambda i})$ is a two-body scalar (Jastrow) correlation between the hyperon and a single nucleon
and $|\Psi_{A}\rangle$ is the correlated wave function describing the remaining $A$ nucleons.
In our calculation this function is defined as:
\begin{equation}
\label{psin}
	|\Psi_{A}\rangle = \left( \prod_{j>i=1}^A f_N(r_{ij}) \right) \Phi_{N,Z}(1,\ldots,A) \,,
\end{equation}
where $\Phi_{N,Z}$ is the Slater determinant of a set of single particle wave functions of $N$ neutrons and $Z$ protons.
Obviously, $A=Z+N$.
\begin{table}[t]
\caption{\label{tabone} Hyperon separation energy $B(\Lambda)$ as computed
by AFDMC simulations for a set of $\Lambda$-hypernuclei. Results
are reported for Hamiltonians with a two body YN interaction, as
described in the text, and for a Hamiltonian including a three-body YNN
interaction. Experimental estimates are from Ref. \cite{Pile:1991}}

\begin{center}
\lineup
\begin{tabular}{lcccccc}
\br                              
&YN only&YN+YNN&E(no $\Lambda$)&$B(\Lambda)_{YN}$&$B(\Lambda)_{YN+YNN}$&$B(\Lambda)_{exp}$\\
\mr
$_{\Lambda}^{5}$He&-33. 4(1)&-30.4(2)&-26.81(8)&6.6(1)&3.6(2)&3.12\cr
$_{\Lambda}^{7}$He&-&-29.5(3)&-25.05(6)&-&4.46&5.23\cr
$_{\Lambda}^{9}$He&-&-28.3(4)&-24.03(8)&-&4.29&?\cr
$_{\Lambda}^{17}$O&-138.0(4)&-117.3(8)&-105.4(1)&32.6(9)&12.0(9)&(13.5)\cr
$_{\Lambda}^{18}$O&-&-118(1)&-104.5(1)&-&13(1)&?\cr
$_{\Lambda}^{41}$Ca&-326(2)&-293(1)&-279.4(5)&47(1)&13(2)&?\cr
\br
\end{tabular}
\end{center}
\end{table}

The AFDMC calculations have been performed for different values of the parameters characterizing the hyperon-nucleon interaction. Despite the estimates of the $\Lambda$ separation energies are still rather noisy, it was possible to determine a set of values of the potential parameters by which the experimental data are roughly reproduced at the same time for $^{5}_{\Lambda}$He and $^{17}_{\Lambda}$O. With this set of parameters we gave a preliminary estimate of the separation energies by both including and excluding the YNN term in the Hamiltonian. Notice that we are making the strong assumption that the separation energy, being the difference of two terms, is not strongly influenced by the quality of the nucleon-nucleon Hamiltonian that we employ in our calculations.

In Table 1 we report the results obtained in our simulations. As it can be seen, the first indication is that using a YN Hamiltonian only, the separation energy tends to be overestimated. On the other hand, the YNN term seems to give an overall repulsive contribution. This result would confirm the fact that the inclusion of three-body forces is necessary to correctly discuss the onset of hyperons in infinite matter, and their contribution to correcting quantities like the mass-radius relation, which are presently the subject of an open discussion.

\section*{Acknowledgments}
The authors from University of Trento are also at LISC, Interdisciplinary Laboratory for Computational Science, a joint venture with the Bruno Kessler Foundation. Part of this work was funded by the PAT/INFN \emph{Aurora} project, and by  a ISCRA cat. B computing grant. We acknowledge J.C. Miller, B. Gibson and J. Carlson for stimulating discussions and contributions to this work. 
The work of S.G. is supported by the U.S. Department of Energy, 
Office of Nuclear Physics, under contracts DE-FC02-07ER41457 (UNEDF SciDAC), and
DE-AC52-06NA25396.

\section*{References}
\bibliographystyle{iopart-num}
\bibliography{eost}
\end{document}